\documentclass[reprint,superscriptaddress,amsmath,amssymb]{revtex4-2}
\usepackage{graphicx,dcolumn,bm,xcolor,mathptmx,etoolbox,blindtext,upgreek,float,nicefrac}
\definecolor{aipblue}{RGB}{46, 169, 224}
\usepackage[colorlinks=true,citecolor=aipblue,urlcolor=aipblue,linkcolor=aipblue]{hyperref}
\usepackage[utf8]{inputenc}
\usepackage[T1]{fontenc}
\setcitestyle{super}
\begin{document}
\title{Synthesis of thin film infinite-layer nickelates by atomic hydrogen reduction: clarifying the role of the capping layer}
\author{C. T. Parzyck}
  \affiliation{Laboratory of Atomic and Solid State Physics, Department of Physics, Cornell University, Ithaca, NY 14853, USA}
\author{V. Anil}
  \affiliation{Laboratory of Atomic and Solid State Physics, Department of Physics, Cornell University, Ithaca, NY 14853, USA}
\author{Y. Wu}
  \affiliation{Laboratory of Atomic and Solid State Physics, Department of Physics, Cornell University, Ithaca, NY 14853, USA}
\author{B. H. Goodge}
  \affiliation{School of Applied and Engineering Physics, Cornell University, Ithaca, NY 14853, USA}
  \affiliation{Kavli Institute at Cornell for Nanoscale Science, Cornell University, Ithaca, NY 14853, USA}
\author{M. Roddy}
  \affiliation{Laboratory of Atomic and Solid State Physics, Department of Physics, Cornell University, Ithaca, NY 14853, USA}
\author{L. F. Kourkoutis}
  \affiliation{School of Applied and Engineering Physics, Cornell University, Ithaca, NY 14853, USA}
  \affiliation{Kavli Institute at Cornell for Nanoscale Science, Cornell University, Ithaca, NY 14853, USA}  
\author{D. G. Schlom}
  \affiliation{Kavli Institute at Cornell for Nanoscale Science, Cornell University, Ithaca, NY 14853, USA}
  \affiliation{Department of Materials Science and Engineering, Cornell University, Ithaca, NY 14853, USA}
  \affiliation{Leibniz-Institut f{\"u}r Kristallz{\"u}chtung, Max-Born-Stra{\ss}e 2, 12489 Berlin, Germany}
\author{K. M. Shen}
  \affiliation{Laboratory of Atomic and Solid State Physics, Department of Physics, Cornell University, Ithaca, NY 14853, USA}
  \affiliation{Kavli Institute at Cornell for Nanoscale Science, Cornell University, Ithaca, NY 14853, USA}
  \affiliation{Institut de Ci{\`e}ncia de Materials de Barcelona (ICMAB-CSIC), Campus UAB Bellaterra 08193, Spain}

\begin{abstract}
  We present an integrated procedure for the synthesis of infinite-layer nickelates using molecular-beam epitaxy with gas-phase reduction by atomic hydrogen.  We first discuss challenges in the growth and characterization of perovskite NdNiO$_3$/SrTiO$_3$, arising from post growth crack formation in stoichiometric films.  We then detail a procedure for fully reducing NdNiO$_3$ films to the infinite-layer phase, NdNiO$_2$, using atomic hydrogen; the resulting films display excellent structural quality, smooth surfaces, and lower residual resistivities than films reduced by other methods. {We utilize the \textit{in situ} nature of this technique to investigate of the role that SrTiO$_3$ capping layers play in the reduction process, illustrating their importance in preventing the formation of secondary phases at the exposed nickelate surface. A comparative bulk- and surface-sensitive study indicates formation of a polycrystalline crust on the film surface serves to limit the reduction process.}
\end{abstract}
\date{\today} 
\maketitle

Over the past four years there has been a renewed interest in the synthesis of the rare-earth perovskite and infinite-layer (IL) nickelates, spurred on by the discovery of superconductivity in the IL nickelates. \cite{Li2019b,Osada2020,Osada2021,Zeng2021}  Owing to challenging synthesis conditions for nickelate single crystals\cite{Zhang2017a,Zheng2019} and their subsequent reduction to the IL phase, \cite{Puphal2021,Puphal2022a,Huo2022,Puphal2023b,Wu2023} much of this work has focused on thin films; to date, superconductivity has not yet been observed in bulk, reduced nickelates.  Correspondingly, there has been a recent renaissance in the study of perovskite nickelates, their metal-to-insulator transition (MIT),\cite{Dominguez2020} and accompanying antiferromagnetism. \cite{Song2023,Dominguez2022}  Nonetheless, the growth and reduction of NdNiO$_3$ thin films are not without their own challenges.  Non-stoichiometry \cite{Breckenfeld2014a,Preziosi2017,Yamanaka2019a,Li2021} and strain\cite{Heo2017,Hauser2015,FerencSegedin2023} both substantially influence the growth and properties of the precursor phase, as well as the quality of subsequently reduced films.\cite{Lee2020}  Furthermore, the typical calcium hydride reduction process employed to form the IL phase typically reduces sample crystallinity, with weaker diffraction peaks observed in reduced films than in their precursors.  This degradation is at least partially attributable to the instability of Ni$^{1+}$ in the IL phase, which decomposes into the thermodynamically preferred $2+$ state at elevated temperatures.\cite{Hayward1999,Hayward2003a}  As a result, long reaction times (typically hours for films and days in bulk) induce a competition between reduction and decomposition during the low-temperature second step.

\begin{figure}\centering
  \resizebox{\columnwidth}{!}{  
  \includegraphics{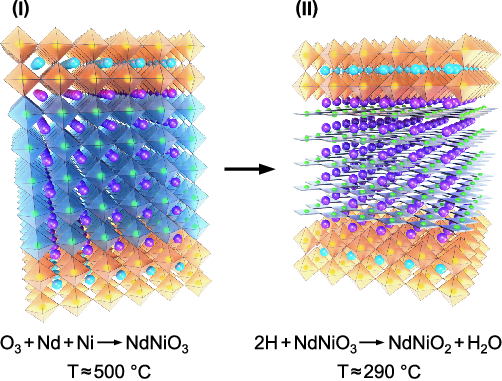}}
  \caption{\label{fig:schematic} Schematic of the two-step \textit{in situ} synthesis of NdNiO$_2$.  (I) First the perovskite NdNiO$_3$ film is grown using distilled ozone and molecular beams of Nd and Ni.  (II) Second, the temperature is lowered and atomic hydrogen is flowed across the sample surface, scavenging oxygen from the film and topotactically reducing it to NdNiO$_2$. H$_2$O generated from this reaction is removed from the system with a high-vacuum pump.}
\end{figure}

In this article we discuss the \textit{in situ} synthesis of NdNiO$_2$ thin films using a combined molecular-beam epitaxy and atomic hydrogen reduction system. We begin with a description of some obstacles to the growth of thin films of NdNiO$_3$ under high tensile strain ($\epsilon = 2.6$\%) on (001) SrTiO$_3$ substrates, including the confounding factors of non-stoichiometry and strain relaxation on the transport properties of precursor films.  We identify a strong aging effect in the resistivity of stoichiometric films and, using a combination of x-ray diffraction (XRD) and high-resolution scanning transmission electron microscopy (STEM), demonstrate it results from post-growth strain relaxation via the formation of microscopic cracks.  After identifying suitable conditions for the growth of the perovskite, we introduce a novel procedure for reduction to the IL phase using an atomic-hydrogen-beam source.\cite{Tschersich1998,Tschersich2008} We show that topotactic reduction of NdNiO$_3$ to NdNiO$_2$ utilizing atomic hydrogen can be accomplished in short periods ($<20$ minutes) and without the need to remove the films from vacuum. Structural and electrical transport measurements indicate a high degree of crystallinity and low surface roughness, establishing this method as a viable alternative to traditional direct contact reduction procedures. { The amenability of this technique to integration with surface sensitive probes is demonstrated by studying the reduction process using reflection high-energy electron diffraction (RHEED). The ability to perform surface-sensitive diffraction allows us to investigate the importance of SrTiO$_3$ capping layers on the reduction of nickelate films, illustrating that uncapped films suffer degradation of the film surface during the reduction.  Furthermore, comparison between bulk and surface sensitive measurements indicate that the formation of this secondary-phase `crust' strongly impacts the reduction process; the same reduction conditions which produce the IL phase in capped samples are insufficient to significantly reduce comparable uncapped samples. The fact that the addition of a single SrTiO$_3$ cap layer to a 20 u.c. thick film drastically changes the reduction rate of the whole film suggests the primary importance of the capping layer is, counterintuitively, to facilitate oxygen removal from the system.}

\section{Synthesis of the Perovskite Precursor}
Historically, nickelate thin films have been synthesized using a wide variety of techniques including pulsed-laser deposition, \cite{Xiang2013} off-axis magnetron sputtering, \cite{Dominguez2020} and molecular-beam epitaxy (MBE); \cite{King2014} in all cases the perovskite exhibits a strong sensitivity to composition, both in terms of the cation ratio\cite{Breckenfeld2014a,Preziosi2017,Yamanaka2019a} and oxygen non-stoichiometry. \cite{Hauser2015,Heo2017} We utilize reactive oxide MBE with distilled ozone ($\sim 80$\%) as the oxidizing agent and a shuttered growth strategy with initial shutter times calibrated using RHEED oscillations measurements of NiO and Nd$_2$O$_3$\cite{Sun2022} and refined using the method described by Li et al.\cite{Li2021} The film composition is inferred from the out-of-plane lattice constant, which has a measurable dependence on the cation ratio to the percent level (\textit{c.f.} supplementary materials).  Following this procedure we obtain a minimal out-of-plane lattice constant of 3.736 \AA, from Nelson-Riley analysis,\cite{Nelson1945} and fully coherent films with sharp MITs.  It has been shown previously that the strength and sharpness of the MIT are strongly influenced by the film non-stoichiometry;\cite{Preziosi2017,Yamanaka2019a,Pan2022a} in our films, we do observe a dependence of the MIT sharpness on the cation ratio, but do not realize a complete suppression of the MIT in the neodymium rich limit (as observed in films under compressive strain \cite{Preziosi2017,Pan2022a}).  Comparison of the film resistivity and MIT transition temperature show monotonic dependence on the lattice constant, further evidencing it as a good metric for the film non-stoichiometry.

\begin{figure}\centering
  \resizebox{\columnwidth}{!}{
  \includegraphics{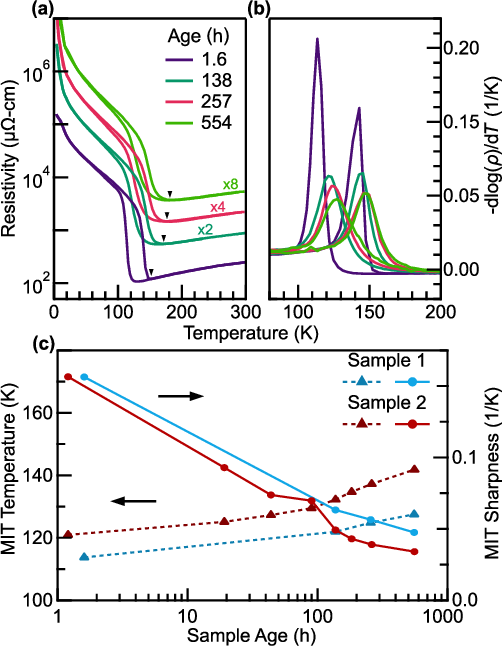}}
  \caption[Aging effects on the resistivity of 20 u.c. thick NdNiO$_3$/SrTiO$_3$]{\label{fig:transportsynt} Aging effects on the electrical transport properties of 20 u.c. thick NdNiO$_3$ films on SrTiO$_3$.  (a) Resistivity of a 20 u.c. film measured repeatedly after its initial growth; traces have been offset for clarity. (b) Logarithmic derivative of the data from panel (a). (c) Transition sharpness and temperature, the maximum value and location of the logarithmic derivative of the resistivity, respectively, for two nominally identical samples measured at different time intervals following growth.}
\end{figure}

\subsection{Resistivity of the Perovskite}
Having tuned the cation ratio to nearly 1-to-1, we investigate the electrical properties of the resulting films.  When the resistivity of a 20 pseudocubic (pc) unit cell (u.c.) film is measured within two hours following growth, \hyperref[fig:transportsynt]{Figs. \ref*{fig:transportsynt}(a)},  the MIT is sharp, with an upturn in the cooling curve at $\sim 140$~K followed by a hysteretic return in the warming curve at $\sim 159$~K.  When the measurement is repeated, however, $\sim 6$~ days later the hysteresis loop is noticeably broadened and these two temperatures are nearly coincident at an elevated value of $\sim 170$~K.  Subsequent measurements show further widening of the transition, an increase of the upturn temperature to $> 180$~K, and an increase in the room temperature resistivity.  The maximum value of the logarithmic derivative of the resistivity, $(-d\log(\rho)/dT)$, gives a measure of the MIT sharpness, and its position provides an even-handed measurement of the transition temperature.  Corresponding curves of the logarithmic derivatives and extracted values are shown in \hyperref[fig:transportsynt]{Figs. \ref*{fig:transportsynt}(b)} and \hyperref[fig:transportsynt]{\ref*{fig:transportsynt}(c)};  the MIT sharpness (temperature) monotonically decreases (increases), roughly logarithmically with time.  A second, nominally identical, sample was measured over the same time period but at a greater frequency and showed identical trends in the sharpness and transition temperatures, verifying that the observed effect is time-dependent rather than a byproduct of thermal cycling.  While the degradation of films under such high tensile strain is not wholly unexpected, the substantial changes in the electrical properties over the course of a single day could confound (epitaxial) strain studies of the MIT in the perovskite nickelates.

\begin{figure*}\centering
  \resizebox{0.9\textwidth}{!}{\includegraphics{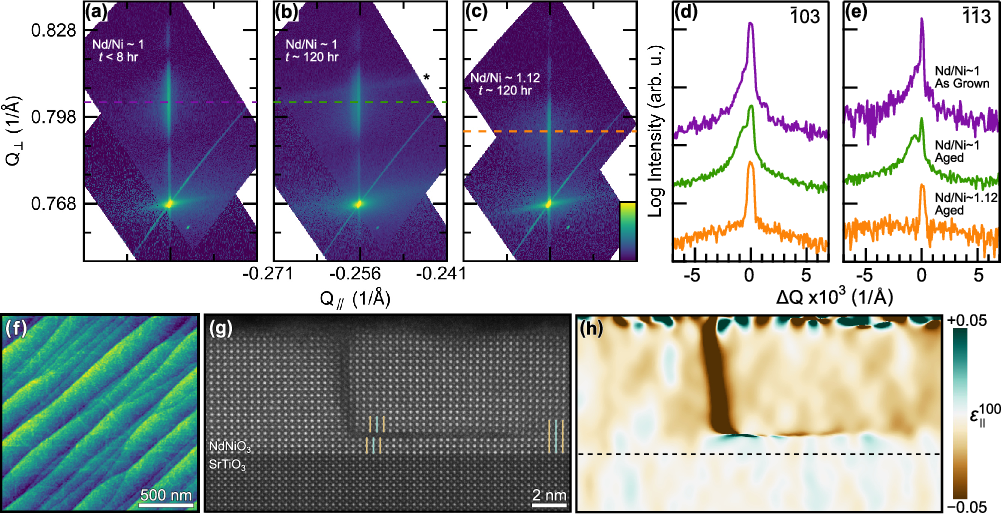}}
  \caption{\label{fig:structure} Aging effects on the structure of 20 u.c. thick NdNiO$_3$ films on SrTiO$_3$. (a) Reciprocal space map about the $\overline{1}03$ peaks of a stoichiometric film measured at a time $t<8$ hours after growth.  (b) Measurement of the same sample \textasciitilde 120 h later; (*) polycrystalline ring from residual silver paint on sample edge. (c) Measurement of a non-stoichiometric film (Nd/Ni$\approx 1.12$) 5 days following growth. (d) Line cuts from panels (a-c) at the peak maximum $\pm 2\times 10^{-3}$~\AA$^{-1}$, traces offset for clarity. (e) Line cuts from equivalent positions on maps about the $(\overline{1}\overline{1}3)_{\textrm{pc}}$ peaks of the same films. (f) AFM image of an aged (> 1 week) NdNiO$_3$ sample showing striations. (g) High-angle annular dark-field image of a crack and resulting delamination in a 3-day-old sample; lines highlight columns of Nd atoms, which should be continuous. (h) Map of in-plane compressive strain in (g) highlighting local changes in the lattice fringe spacing.}
\end{figure*}

\subsection{Post-growth Structural Relaxation}
In \hyperref[fig:structure]{Fig. \ref*{fig:structure}}, we show that the observed changes in resistivity are attributable to post growth strain relaxation by way of crack formation.  A reciprocal space map (RSM) about the $(\overline{1}03)_{\textrm{pc}}$ reflection of a stoichiometric film, measured immediately following growth, shows a sharp peak aligned with the substrate; upon repetition of the measurement 5 days later, a significant shoulder is apparent.  When a purposefully neodymium rich film (Nd:Ni $\sim 1.12$) is measured 5 days following growth, however, no shoulder is observed, indicating the film remains coherent.  This, and equivalent data about $(\overline{1}\overline{1}3)_{\textrm{pc}}$ summarized in \hyperref[fig:structure]{Figs. \ref*{fig:structure}(d)} and \hyperref[fig:structure]{\ref*{fig:structure}(e)}, are consistent with fully coherent as-grown films, regardless of the cation non-stoichiometry. Nonetheless, films with Nd:Ni $\sim 1$~ do not maintain coherence on the hour timescale. Instead, they form regions with a smaller in-plane lattice parameter: $a_{\textrm{pc}} \sim 3.891(5)$ -- intermediate to the substrate (3.905~ \AA) and bulk\cite{Klein2021} (3.807~\AA) parameters.  Partial relaxation towards the bulk-like structure explains the time dependent increase in the MIT temperature, towards that of the bulk compound. The fact that neodymium-rich films do not form relaxed regions suggests the presence of Ruddlesden-Popper (RP) faults, identified in Nd rich films,\cite{Lee2020} contribute to the accommodation of both off-stoichiometry as well as high tensile strain.

Real-space probes including atomic force microscopy (AFM) and STEM reveal relaxation in the film is accomplished by crack formation and subsequent partial delamination.  AFM of an aged, stoichiometric film, \hyperref[fig:structure]{Fig. \ref*{fig:structure}(f)},  shows terraces decorated with raised striations running along and across the step edges, indicative of cracking and subsequent delamination, as observed prior in perovskite oxides under biaxial tensile strain. \cite{Biegalski2008} This partial delamination is directly visible by atomic-resolution HAADF-STEM imaging in \hyperref[fig:structure]{Fig. \ref*{fig:structure}(g)} where the top surface of the film is slightly raised on one side of the crack's vertical extension.  In addition to the vertical lifting of the atomic layers, in-plane relaxation above the crack can be identified by Nd atomic columns, which are offset above and below the horizontal extension (cyan and yellow lines) and by a small effective compression of the in-plane lattice fringe spacing\cite{Goodge2022b} (light brown) in 
\hyperref[fig:structure]{Fig. \ref*{fig:structure}(h)}.  In addition to the aforementioned time dependence in electrical transport caused by partial relaxation, the formation of extended defects may present a challenge in the reproducibility of reduction experiments. If the time between growth and reduction is not controlled, as the off-stoichiometry of the parent phase is reduced and frequency of RP faults diminished, the quality of the corresponding reduced films may become difficult to predict.

\section{Reduction to infinite-layer N\texorpdfstring{\MakeLowercase{d}}{}N\texorpdfstring{\MakeLowercase{i}}{}O$_2$}
\begin{figure}\centering
  \resizebox{\columnwidth}{!}{\includegraphics{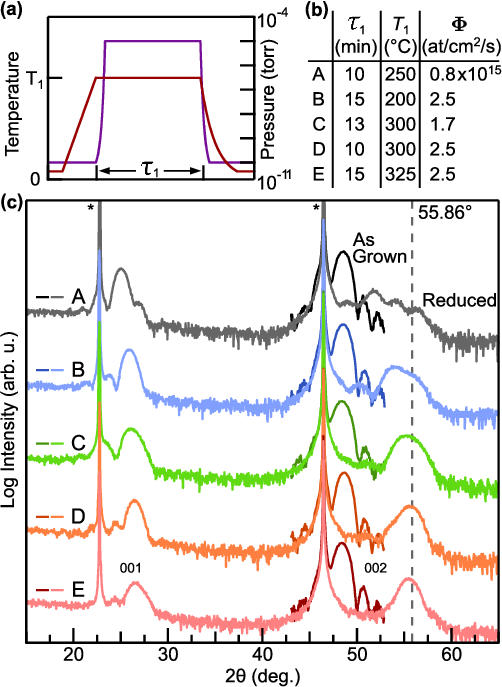}}
  \caption{\label{fig:reduction} Single temperature topotactic reduction of NdNiO$_3$/SrTiO$_3$ with atomic hydrogen. (a) Schematic of reduction procedure at a single temperature $T_1$ for a duration $\tau_1$.  (b) Time, temperature, and flux conditions for reduction of samples shown in (c). (c) X-ray diffraction of a series of samples before (dark) and  after (light) reduction, offset for clarity. All samples were capped with 1-3 layers of SrTiO$_3$. (*) indicates reflections arising from the SrTiO$_3$ substrate.}
\end{figure}

Having established conditions for synthesis of the perovskite, we introduce a novel reduction procedure to transform it to the infinite-layer phase.  Traditionally, low-temperature topotactic reduction of nickelates utilizes alkali or alkaline earth hydride powders (CaH$_2$, NaH) as reducing agents.  This process was introduced in studies of intermixed polycrystalline samples, \cite{Crespin1983a,Hayward1999,Hayward2003a,Crespin2005} later applied to films,  \cite{Kawai2009,Ikeda2016a,Onozuka2016} and most recently has been refined for the production of superconducting alkaline-earth (\textit{AE}) doped rare-earth (\textit{RE}) nickelates, \textit{(RE,AE)}NiO$_2$.\cite{Li2020a,Zeng2020,Osada2020a,Osada2020,Li2019b,Lee2020} Despite the success of this method there are a few drawbacks. Primarily, it is limited to \textit{ex situ} use -- the film needs to be removed from the growth chamber and placed in an environment with the powder, annealed, and then the reaction products removed.  Often, the sample is embedded in the powder \cite{Krieger2023} (though there have been reports of non-contact reduction\cite{Ikeda2014,Gao2021a,Puphal2021}), which is not amenable to surface sensitive measurements, including scanning tunneling microscopy and photoemission spectroscopy.  These techniques, fundamental to understanding the cuprates, remain uncommon in the infinite-layer nickelates.\cite{Gu2020,Chen2022,Wang2023}  Additionally, long reduction times (typically hours for thin films and days for single crystals) may be detrimental to preservation of crystallinity as bulk studies indicate extended exposure to temperatures exceeding 200 $^{\circ}$C results in decomposition or disproportionation of the metastable structure.\cite{Hayward1999,Hayward2003a,Puphal2022a}  We note that there has been a recent push to identify new reduction methods for nickelates, and reduction by application of an aluminum getter layer to the film surface has been recently demonstrated. \cite{Wei2023,Wei2023a}  Similar to the recently identified solid-state aluminum reduction technique, the method described here can be achieved in the same vacuum system as the sample growth. 

The use of atomic hydrogen as a powerful reducing agent has been recognized previously; \cite{Bergh1965} however, to date its use has primarily been consigned to surface etching \cite{Gates1989} and cleaning, \cite{Bell1998,Nishiyama2005} with some work on the reduction of binary oxides, \cite{Huang2007,Knudsen2010,Shahed2014a} or generation of oxygen vacancies in more complex oxides.\cite{Takahashi2004}  We find that exposure to an atomic hydrogen flux of $\sim 2\times 10^{15}$~at/cm$^2$/sec for $<15$ minutes at sample temperatures between 250 and 300 $^{\circ}$C is sufficient to transform $\sim 7.5$~nm thick perovskite films to the IL phase, a much shorter exposure compared to traditional CaH$_2$ annealing.\cite{Meng2023} We further demonstrate that this procedure produces IL films with high structural quality, smooth surfaces, and favorable electrical characteristics. Finally, we leverage the compatibility of this technique with \textit{in situ} probes, namely RHEED, to study the effects of atomic and molecular hydrogen exposure on the film surface and the resulting impact on the reduction process.

\subsection{Single Temperature Reduction}
Following growth, 20 u.c. thick nickelate films are capped with 2-3 unit cells of SrTiO$_3$ and transferred, in ultra-high vacuum, to a connected chamber ($P_{\textrm{base}}<1\times 10^{-10}$~torr) with a heated sample stage.  There, atomic hydrogen is generated from a thermal source \cite{Tschersich1998,Tschersich2008} by passing molecular H$_2$ through a heated tungsten capillary ($>1900$~$^{\circ}$C) located 9 cm from the sample. It disassociates (with an efficiency, $\alpha$, between 50 and 80\%) \cite{Zheng2006} before interacting with the sample surface, scavenging O$_2$ from the film, and forming H$_2$O, which is removed via a turbomolecular pump.  A schematic of a single-temperature reduction program, with typical parameters, is shown in \hyperref[fig:reduction]{Figs. \ref*{fig:reduction}(a)} and \hyperref[fig:reduction]{\ref*{fig:reduction}(b)}; lab-based XRD measurements before and after reduction are detailed in \hyperref[fig:reduction]{Fig. \ref*{fig:reduction}(c)}.  At low reaction times, temperatures, and fluxes (samples A, B) the perovskite transforms partially to the IL phase with a pair of additional intermediate peaks visible at $2\theta=51.8^{\circ}$ ($c\sim3.53$~\AA) and $2\theta=54.0^{\circ}$ ($c\sim3.39$~\AA),  attributable to oxygen-deficient perovskite phases with partially collapsed structures.  As the reaction temperature is increased from 250 to $\sim 300$~$^{\circ}$C and the flux from $<1$ to $2.5\times 10^{15}$~at/cm$^2$/sec (by varying the capillary temperature and H$_2$ flow rate) the intermediate phase peaks are suppressed in favor of a single peak above $2\theta=55^{\circ}$, indicative of the IL phase (Sample E).  Eliminating these oxygen-rich phases from NdNiO$_2$ samples is of paramount importance to studies targeted to understanding of the parent state from which superconductivity emerges.  Excess interstitial oxygen tends to form ordered structures in all members of the \textit{RE}NiO$_2$, \textit{RE}=La, Pr, Nd, family. \cite{Gonzalez1989,Moriga1994a,Sayagues1994,Alonso1995,Moriga2002,Wang2020,Wu2023,Raji2023}  Inclusions of these phases may serve to confound resonant scattering studies, which are highly sensitive to periodic valence modulations of the nickel. \cite{Parzyck2023}

\subsection{Three-Step Reduction}
\begin{figure*}\centering
  \resizebox{0.9\textwidth}{!}{\includegraphics{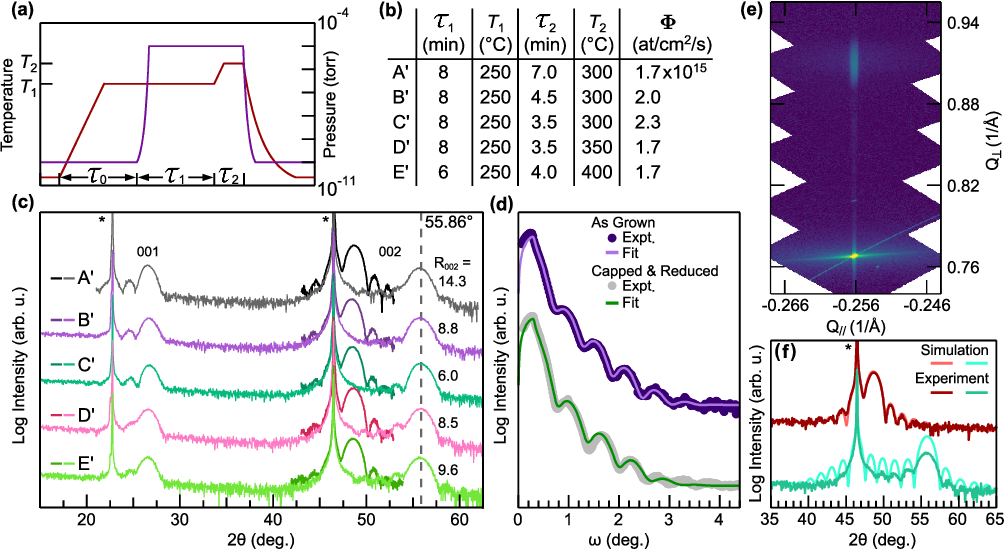}}
  \caption{\label{fig:optimal} Structural characterization of optimally reduced NdNiO$_2$/SrTiO$_3$. (a) schematic of the three-step reduction procedure. (b) Parameters used in the reduction of samples in (c); in all pictured cases an initial equilibration time of $\tau_0=8$ minutes was used. (c) XRD of samples before (dark) and after (light) reduction. The ratio of the 002 peak intensity before and after reduction ($R_{002}=I(2\theta=48.71^{\circ})/I'(2\theta=55.86^{\circ})$) is shown as a metric of the conversion efficiency. (d) X-ray reflectivity of the same sample immediately after growth (black) and after capping with 2 u.c. of SrTiO$_3$ and reduction using the optimal conditions of C'. (e) Reciprocal space map about the $\overline{1}03$ peaks of SrTiO$_3$ and NdNiO$_2$ for a sample reduced under the optimized conditions. (f) XRD, about the 002 reflection, of representative, optimized NdNiO$_3$ and NdNiO$_2$ films with simulated diffraction patterns overlaid. (*) indicates reflections arising from the SrTiO$_3$ substrate.}
\end{figure*}

While the single temperature reduction procedure is shown to effectively produce the IL phase, the resulting films have reduced crystallinity compared to the perovskite, evidenced by reduction in the (002) Bragg peak intensity. We quantify the conversion efficiency from the perovskite to the IL phase by the ratio of the (002) peak intensity before ($I$) and after ($I'$) reduction: $R_{002}=I(2\theta=48.71^{\circ})/I'(2\theta=55.86^{\circ})$.  Owing to changes in the structure factor, some decrease in intensity is expected: the optimal (\textit{i.e.} minimal) value of $R_{002}$ is calculated to be 1.58 using a dynamical diffraction model.\cite{Kriegner2013} For reference, the value of $R_{002}$, for sample E in \hyperref[fig:reduction]{Fig. \ref*{fig:reduction}(b)} is 13.1.  This discrepancy highlights that although the reduction may proceed across the entire thickness of the film (evidenced by the Scherrer thickness from the 002 peak) the coherent crystalline volume of the reduced film is reduced from that of the perovskite.

To further improve the crystallinity, we perform a three-step reduction procedure utilizing the independent control of the sample temperature and reducing atmosphere afforded in this setup.  Outlined in \hyperref[fig:optimal]{Fig. \ref*{fig:optimal}(a)} and \hyperref[fig:optimal]{\ref*{fig:optimal}(b)}, the procedure consists of: (1) heating close to the reduction temperature ($\sim$250~$^{\circ}$C) in vacuum for 8 minutes to achieve thermal equilibrium;  (2) the hydrogen flow is turned on and the sample exposed to the atomic H beam for 6-8 minutes as the flow rate stabilizes;  (3) the sample is quickly heated to the final temperature for a short period, typically 3-5 minutes, before terminating the hydrogen flow and cooling the sample.  As illustrated in the corresponding XRD measurements, \hyperref[fig:optimal]{Fig. \ref*{fig:optimal}(c)}, using similar parameters to those in \hyperref[fig:reduction]{Fig. \ref*{fig:reduction}} results in similar values of $R_{002}$ (Sample A'). Lowering the reduction time while increasing the temperature and flux appears to improve the conversion efficiency; with full reduction achievable in only 10-12 minutes of total exposure time (Sample C'- E'), with an optimized value of $R_{002}\approx 6$~in sample C'.  The precise value of the high-temperature step does not appear to strongly influence $R_{002}$; this is possibly attributable to the short duration of the final step, during which the sample surface may not fully equilibrate to the heater.  Although the conditions of sample C' provide the best conversion efficiency, \hyperref[fig:optimal]{Fig. \ref*{fig:optimal}(b)} illustrates that the IL phase can be isolated over a reasonable window of times, temperatures, and fluxes.

Further structural characterization is outlined in \hyperref[fig:optimal]{Fig. \ref*{fig:optimal}(d)-\ref*{fig:optimal}(f)}.  X-ray reflectivity (XRR) of a film is measured immediately after growth and then again following capping with 2 u.c. of SrTiO$_3$ and reduction.  The thickness and roughness of the NdNiO$_x$ slab obtained from a dynamical fit of the XRR\cite{Kriegner2013} are $\ell_{PER}=7.31$~nm, $\sigma _{PER}=0.5$~nm before the reduction and $\ell_{IL}=6.20$~nm, $\sigma _{IL}=1.9$~nm after. The thickness ratio of $\ell_{IL}/\ell_{PER}=0.85$ matches the ratio of the $c$-axis lattice constants, 0.88 -- consistent with a full collapse of the structure without substantial increase in the sample roughness.  An RSM about the $\overline{1}03$ peaks of SrTiO$_3$ and NdNiO$_2$ shows that coherence of the film is maintained through the reduction process. In a 00L scan of a 3 u.c. SrTiO$_3$ capped film (supplemental materials) all of the 001 through 004 reflections are visible in the IL film, and an out-of-plane lattice parameter of $c=3.286$~\AA~is calculated from Nelson-Riley analysis,\cite{Nelson1945} in good agreement with previous measurements of chemically reduced films.\cite{Li2020a,Goodge2022a}  A simulation of unreduced NdNiO$_3$/SrTiO$_3$ is fit to the XRD data in \hyperref[fig:optimal]{Fig. \ref*{fig:optimal}(f)} to extract the setup parameters, which are then used to simulate the XRD for an NdNiO$_2$ film (changing only the structure from NdNiO$_3$ to NdNiO$_2$).  Comparing to a capped/reduced film, the observed pattern of Pendell\"{o}sung fringes (and the Scherrer thickness) are correctly reproduced in the simulation, including the fringe spacing and the asymmetric intensity pattern.  

\subsection{Electrical Transport Measurements}
\begin{figure}\centering
  \resizebox{\columnwidth}{!}{
  \includegraphics{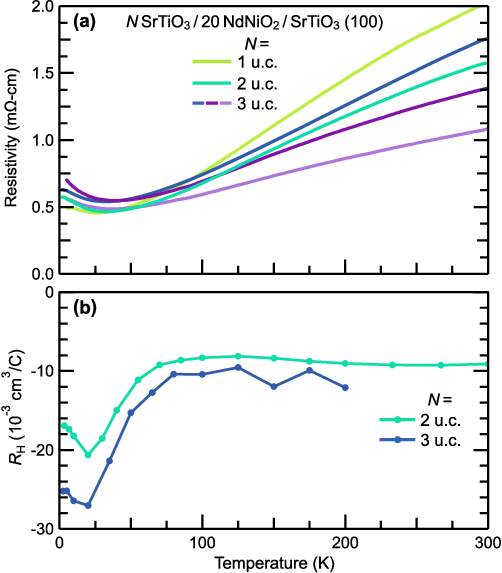}}
  \caption[Resistivity and hall coefficient measurements of topotactically reduced nickelate films]{\label{fig:redtransport} Electrical transport measurements of reduced nickelate films.  (a) Resistivity of representative reduced films with different SrTiO$_3$ cap thicknesses, \textit{N}, reduced using the optimized three-step procedure. (b) Hall measurements of NdNiO$_2$ films capped with 2 or 3 u.c. SrTiO$_3$.}
\end{figure}

The resistivity of a representative set of films reduced using the three-step procedure is shown in \hyperref[fig:redtransport]{Fig. \ref*{fig:redtransport}}.  Transport measurements of films, with varying cap thickness, show residual resistivities between 300 and 600 $\mu\Omega$-cm; this is lower than comparable undoped nickelate films on (001) SrTiO$_3$ \cite{Li2020a,Osada2020a,Wei2023} and comparable to underdoped films reported on (001) LSAT \cite{Wei2023a,Lee2023} prepared by other methods.  All samples show some degree of an upturn at low temperature, between 20 and 35 K.  Hall measurements, detailed in \hyperref[fig:redtransport]{Fig. \ref*{fig:redtransport}(b)}, show a flat temperature dependence of $R_H$ between 75 and 300 K, consistent with prior measurements on undoped and underdoped NdNiO$_2$, \cite{Li2020a,Zeng2020,Ikeda2016a} accompanied by a downturn at low temperature.  The Hall coefficient, $R_H$, in the high temperature limit is around -0.01 cm$^3$/C, which is intermediate to prior measurements of films on SrTiO$_3$, between -0.005 reported by Li et al. \cite{Li2020a} and -0.035 as reported by Zeng et al. \cite{Zeng2020} In contrast to prior measurements, a slight upturn in $R_H$ is visible below 25 K. However, we note that the precise shape of $R_H(T)$ reported in the literature appears to vary considerably between different samples, \cite{Li2020a,Zeng2020} A-site cations, \cite{Osada2020a} and strain states. \cite{Lee2023}

\subsection{Effects of the Capping layer}
\begin{figure*}\centering
  \resizebox{0.9\textwidth}{!}{\includegraphics{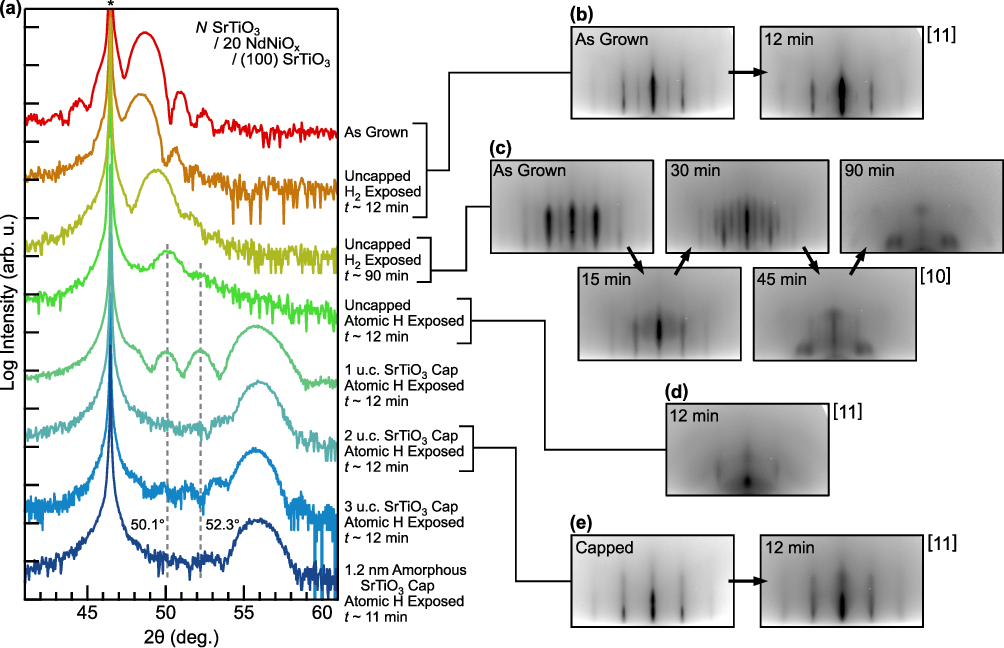}}
  \caption{\label{fig:cappingsynt} Surface- and bulk-sensitive structural measurements of NdNiO$_{3-x}$ films reduced using molecular and atomic hydrogen.  (a) XRD of capped and uncapped nickelate samples exposed to molecular or atomic hydrogen. Dashed lines indicate the position of peaks likely derived from intermediately reduced phases and (*) indicates reflections arising from the SrTiO$_3$ substrate; traces have been vertically offset for clarity. (b) RHEED images, taken with the electron beam along the [110]$_{pc}$ azimuth, of an uncapped perovskite film before and after a short exposure to molecular hydrogen.  (c) RHEED images, taken with the electron beam along the [100]$_{pc}$ azimuth, of an uncapped film progressively exposed for longer periods, up to a total time of 1.5 hours. (d) RHEED images, taken with the electron beam along the [110]$_{pc}$ azimuth, of an uncaped film exposed for a short duration to atomic hydrogen. (e) RHEED images, taken with the electron beam along the [110]$_{pc}$ azimuth, of a film capped with 2 u.c. of SrTiO$_3$ exposed to atomic hydrogen.  All RHEED images use a logarithmic intensity scale.  Samples were reduced under similar conditions with chamber pressures of $\sim 9\times 10^{-5}$~torr and sample temperatures between 300 and 400 $^{\circ}$C. Precise values of the times, temperatures, and fluxes used for each sample are tabulated in the supplementary materials.}
\end{figure*}

{ We conclude with a study of the effects of molecular and atomic hydrogen on film surfaces during reduction.  Earlier reports have speculated on the role of the capping layer during reduction; our experiments reveal that exposure of the nickelate surface to a reducing atmosphere results in the formation of a polycrystalline scale layer on the sample surface. Furthermore, uncapped samples with this scale layer are \textit{not} reduced under the same conditions which are sufficient for reduction of films capped with caps as thin as 1 u.c. of SrTiO$_3$.} In previous sections we have described samples capped with 1-3 u.c. of SrTiO$_3$ and reduced with atomic hydrogen.  Here we describe what happens to films without a capping layer when they are exposed to H and H$_2$.  \hyperref[fig:cappingsynt]{Figure \ref*{fig:cappingsynt}(a)} displays XRD of a series of 20 u.c. thick films (capped and uncapped) following exposure to molecular or atomic hydrogen under similar conditions (chamber pressure and substrate temperature). Corresponding RHEED measurements for selected samples are included in \hyperref[fig:cappingsynt]{Figs. \ref*{fig:cappingsynt}(b)-\ref*{fig:cappingsynt}(e)} and the specific details of each program are provided in the supplemental materials. Exposure of an uncapped sample to molecular hydrogen for a short duration (< 15 min) at a flux of $1.5\times 10^{15}$~ H$_2$/cm$^2$/sec produces no apparent change in the XRD. However, the half-order streaks (associated with rotation of the NiO$_6$ octahedra) are diminished, indicating a change in the surface oxygen stoichiometry.  15 minutes of additional exposure produces $\nicefrac{1}{3}^{\textrm{rd}}$~ order peaks, and with further reduction the disappearance of film peaks and formation of a secondary phase are observed.  This degradation of the surface occurs at timescales short compared to the reduction of the entire film, evidenced by the modest shift in the XRD.  Additionally, the total exposure of the film surface to H$_2$ gas in this experiment is low, limited to $< 10^6$~ Langmuir (L) over 90 minutes, compared to hydride annealing experiments where the sample environment may contain up to 100 mtorr of hydrogen\cite{Zeng2020} (liberated from CaH$_2$ or Ca(OH)$_2$\cite{Ikeda2014,Kobayashi2013}) for several hours ($\sim 10^9$~ L).

This surface degradation also affects oxygen removal during the atomic hydrogen reduction.  When an uncapped film is reduced using the three-step process with nominally optimized conditions (12 minutes of total exposure) we observe a clear degradation of the RHEED pattern: the film peak intensity is reduced, background intensity increased, and faint polycrystalline rings appear.  Surprisingly,  this exposure does not produce an IL film; instead, a pair of intermediate phase peaks are observed at $2\theta$~ values of 50.1$^{\circ}$ and 52.3$^{\circ}$.  The addition of a capping layer greatly changes the process: using nearly identical time, temperature, and flux conditions, films with 1-3 layers of SrTiO$_3$ \textit{are} reduced to the IL phase.  In the 1 u.c. thick SrTiO$_3$ capped film, additional peaks are visible at positions not associated with Pendell\"{o}sung fringes for this thickness film.  These peaks, matching the positions of those in the uncapped film, suggest trace intermediate phases may remain; which are absent in the films capped with 2 or 3 u.c. of SrTiO$_3$.  RHEED measurements show the capped films maintain an ordered surface and AFM (supplementary materials) indicates a smooth, terraced surface is maintained, with a terrace-averaged roughness of only 1.2 \AA. The substantial difference in oxygen removal between capped and uncapped films, under the same reduction conditions, highlights the cap's importance in facilitating the reduction process.  

These extremely thin capping layers, approaching a single unit cell, likely do not substantially contribute to maintaining the structural stability across the entirety of the 20 u.c. film; yet they substantially influence the conditions required to reduce the film.  This suggests that the cap plays a role beyond that of a simple structural support and has a substantial impact on how easily oxygen is removed from the film.  This distinction can be further explored by comparing the reduction of a sample capped with 3 u.c. of crystalline SrTiO$_3$ with one capped with 1.2 nm of amorphous SrTiO$_3$.  We observe (\textit{c.f.} \hyperref[fig:cappingsynt]{Fig. \ref*{fig:cappingsynt}(a)}) that samples capped with amorphous SrTiO$_3$ are well reduced using the three-step atomic hydrogen reduction procedure, even though the cap does not provide additional epitaxial support to the film.  While films with either crystalline or amorphous SrTiO$_3$~ caps are both reduced under similar conditions, there are some slight differences in the quality of the resulting IL films -- discussed in the supplemental materials -- with amorphous SrTiO$_3$ capped films suffering from reduced crystallinity.  The similarity between samples capped with crystalline and amorphous material, contrasted with the distinct behaviour of uncapped samples, strongly suggests that during the atomic hydrogen reduction the primary function of the cap is to prevent the formation of a nickelate scale layer at the sample surface, rather then providing epitaxial stabilization of the IL phase.  The contrast illustrated here may help to explain observed discrepancies between capped and uncapped IL nickelate films in recent transport\cite{Krieger2023} and scattering\cite{Krieger2022} studies. 

\section{Conclusions}
Here, we have described a novel, flexible synthesis method for IL nickelate films -- one which requires only gas phase reactants and is easily integrable with thin film growth techniques and surface sensitive probes.  Characterization of the resulting films shows this method reproducibly yields IL films with high crystallinity, low resistivity, and flat surfaces. { Utilizing the \textit{in situ} nature of this procedure, we have investigated the effects of the reduction on the film surface to elucidate the impact of capping layers on the reaction. We find that the addition of an ultrathin cap, down to 0.4-1.2 nm, drastically improves the reduction process and inhibits the formation of a polycrystalline scale layer on the sample surface}
Additionally, we have demonstrated the growth of high quality films of NdNiO$_3$ by MBE and illustrated that the structural and electrical characteristics of the highly strained perovskite are unstable on the timescale of hours to days. Diffraction and real-space probes indicate that the sample strain is partially relieved by the rapid formation of cracks soon after growth.  These measurements indicate that care must be taken when investigating the details of the MIT in highly strained nickelates as well unveiling a potentially uncontrolled variable in nickelate reduction experiments.  Topotactically reduced oxides have proven to be a rich, expanding field for the investigation of novel ground states in quantum materials; \cite{Crespin1983a, Hayward1999, Li2019b,Pan2022,Kim2023,Meng2023,Tsujimoto2007,Inoue2008,Hadermann2009,Matsumoto2010,Moon2014,Mairoser2015} we put forth atomic hydrogen reduction as a highly tunable process for synthesizing reduced oxides of high quality.  The vacuum compatibility of this technique both expands the range of experiments which may be performed on reduced films and enables \textit{in situ} and \textit{operando} studies of the reduction process itself.
\section{Methods}
\subsection*{Sample Growth \& Characterization}
Thin films of NdNiO$_3$ were grown on (001)-oriented SrTiO$_3$ substrates using reactive-oxide molecular-beam epitaxy in a Veeco GEN10 system from elemental beams of Nd (Alfa/AESAR, 99.9\%) and Ni (Alfa/AESAR, 99.995\%) with typical fluxes of 0.8-1.2 $\times10^{13}$ at/cm$^2$/sec.  Substrates were etched to prepare a TiO$_2$ terminated surface\cite{Koster1998} and annealed prior to growth at 650 $^{\circ}$C until a clear RHEED pattern was observed. Growths were performed at substrate temperatures between 480 and 500 $^{\circ}$C, measured by optical pyrometry at a wavelength of 1550 nm, in background pressures between 2 and $7\times10^{-6}$ torr of $\sim 80$\% distilled ozone. Following growth, samples were cooled to  $<100$~$^{\circ}$C in a pressure of $>1\times 10^{-6}$ torr of ozone to discourage the formation of oxygen vacancies.  

Structural quality and phase purity were determined using Cu K$\alpha_1$ x-ray diffraction measurements performed on a PANalytical Empyrean X-ray diffractometer. Morphology was assessed using an Asylum Research Cypher environmental atomic force microscope. Electrical transport measurements were performed using both a custom built LHe-cooled four-point probe measurement station and a Quantum Design physical property measurement system utilizing either built-in electronics or a Keithley 6221 current source and 2182A voltmeter in a delta mode configuration. Geometric factors in the resistivity were accounted for using the methods of Miccoli et al.\cite{Miccoli2015}  Hall measurements were performed either over the entire $10\times 10$~mm sample in a square four-point Van der Pauw configuration or using a 6-wire Hall bar geometry on a diced $5\times 10$~mm piece.
\subsection*{Scanning Transmission Electron Microscopy}
Cross-sectional specimens were prepared using the standard focused ion beam (FIB) lift-out process on a Thermo Fisher Scientific G4 UX FIB. HAADF-STEM images were acquired on an aberration-corrected FEI Titan Themis operating at 300 kV with a probe-forming semi-angle of 21.4 mrad and inner (outer) collection angles of 68 (340) mrad.  To avoid possible image distortions from sample drift or other mechanical instabilities, all images presented here are produced by summing a stack of 50 high-frame-rate images aligned by rigid registration. \cite{Savitzky2018} A small-angle shear transformation was applied to all images to correct for subtle non-orthogonality in the STEM scan.  Lattice relaxation in the film was mapped with strain analysis based on a phase lock-in approach.\cite{Goodge2022b}

\section*{Supplementary Material}
See the supplementary material for additional details about the stoichiometric calibration process, further data on the effects of non-stoichiometry and strain relaxation on transport measurements, and additional details about the effects of the reduction process on capped and uncapped films.  
\section*{Acknowledgments}
This work was primarily supported by the National Science Foundation through Grants No. DMR-2104427 and the Platform for the Accelerated Realization, Analysis and Discovery of Interface Materials (PARADIM) under Cooperative Agreement No. DMR-2039380. Additional support was provided by the Air Force Office of Scientific Research (Grant No. FA9550-21-1-0168) and the Gordon and Betty Moore Foundation’s EPiQS Initiative through Grant Nos. GBMF3850 and GBMF9073. The authors acknowledge the use of facilities and instrumentation supported by NSF through the Cornell University Materials Research Science and Engineering Center DMR-1719875.  The FEI Titan Themis 300 was acquired through No. NSF-MRI-1429155, with additional support from Cornell University, the Weill Institute, and the Kavli Institute at Cornell. The Thermo Fisher Helios G4 UX FIB was acquired with support by NSF No. DMR-1539918.  Substrate preparation was performed, in part, at the Cornell NanoScale Facility, a member of the National Nanotechnology Coordinated Infrastructure (NNCI), which is supported by the National Science Foundation (Grant NNCI-2025233). The authors would like to thank Sean Palmer and Steven Button for their assistance in substrate preparation as well as Hari Nair and Jiaxin Sun for their thoughtful discussions and insightful comments.
\section*{Author Declarations}
\subsection*{Conflict of Interest}
The authors have no conflicts to disclose
\subsection*{Author Contributions}
Samples were synthesized, reduced, and characterized by C.T.P, V.A., Y.W., and M.R. under the supervision of D.G.S. and K.M.S.  Electron microscopy was performed by B.H.G. under the supervision of L.F.K.. The manuscript was prepared by C.T.P. and K.M.S. with input from all authors.
\section*{Data Availability}
The data that support the findings of this study are available within the paper and supplementary material. Additional data related to this study are available from the authors upon reasonable request.
\bibliography{Nickelates}

\end{document}


\title{Supplemental Materials for: Synthesis of thin film infinite-layer nickelates by atomic hydrogen reduction: clarifying the role of the capping layer}
\author{C. T. Parzyck}
  \affiliation{Laboratory of Atomic and Solid State Physics, Department of Physics, Cornell University, Ithaca, NY 14853, USA}
\author{V. Anil}
  \affiliation{Laboratory of Atomic and Solid State Physics, Department of Physics, Cornell University, Ithaca, NY 14853, USA}
\author{Y. Wu}
  \affiliation{Laboratory of Atomic and Solid State Physics, Department of Physics, Cornell University, Ithaca, NY 14853, USA}
\author{B. H. Goodge}
  \affiliation{School of Applied and Engineering Physics, Cornell University, Ithaca, NY 14853, USA}
  \affiliation{Kavli Institute at Cornell for Nanoscale Science, Cornell University, Ithaca, NY 14853, USA}
\author{M. Roddy}
  \affiliation{Laboratory of Atomic and Solid State Physics, Department of Physics, Cornell University, Ithaca, NY 14853, USA}
\author{L. F. Kourkoutis}
  \affiliation{School of Applied and Engineering Physics, Cornell University, Ithaca, NY 14853, USA}
  \affiliation{Kavli Institute at Cornell for Nanoscale Science, Cornell University, Ithaca, NY 14853, USA}  
\author{D. G. Schlom}
  \affiliation{Kavli Institute at Cornell for Nanoscale Science, Cornell University, Ithaca, NY 14853, USA}
  \affiliation{Department of Materials Science and Engineering, Cornell University, Ithaca, NY 14853, USA}
  \affiliation{Leibniz-Institut f{\"u}r Kristallz{\"u}chtung, Max-Born-Stra{\ss}e 2, 12489 Berlin, Germany}
\author{K. M. Shen}
  \affiliation{Laboratory of Atomic and Solid State Physics, Department of Physics, Cornell University, Ithaca, NY 14853, USA}
  \affiliation{Kavli Institute at Cornell for Nanoscale Science, Cornell University, Ithaca, NY 14853, USA}
  \affiliation{Institut de Ci{\`e}ncia de Materials de Barcelona (ICMAB-CSIC), Campus UAB Bellaterra 08193, Spain}

\date{\today} 
\maketitle
\tableofcontents
\clearpage

\section{Perovskite Nickelates}

The stoichiometric calibration of nickelate films is of paramount importance to the growth of high-quality NdNiO$_3$.  As the binary oxide constituents, Nd$_2$O$_3$ and NiO have negligible volatility, an absorption-controlled approach in molecular beam epitaxy (MBE) is not possible -- excess deposited material either incorporates or forms precipitates--both of which negatively impact the film quality.  Studies of the pulsed laser deposition (PLD) and MBE growth of nickelates indicate that deviation from a cation ratio of 1:1 has strong effects on the electrical properties of NdNiO$_3$ films\cite{Breckenfeld2014a,Preziosi2017,Yamanaka2019a,Pan2022a} and a strong influence on the quality of subsequently reduced films -- particularly the presence of Ruddlesden-Popper (RP) intergrowths \cite{Lee2020,Li2021,Lee2023}.  In PLD growth, careful control of the film stoichiometry can be achieved through control of the laser fluence and source material (by repolishing the target surface to combat laser-induced changes \cite{Preziosi2017}, for instance).  In MBE, on the other hand, the key is to  establish the elemental source fluxes with a high degree of accuracy.  We find that the typical accuracy of a quartz crystal microbalance (QCM), 10-15\%, is not sufficient to reproducibly grow high-quality nickelate films; we follow instead a two-step calibration procedure detailed here.

\begin{figure}
  \resizebox{14.5 cm}{!}{
  \includegraphics{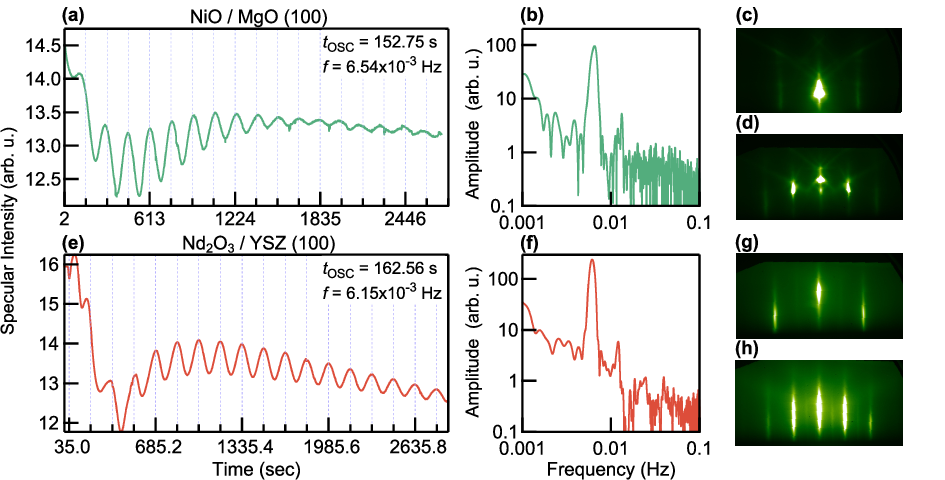}}
  \caption{\label{S:OSC} RHEED Oscillations of the binary oxides NiO/MgO (100) and Nd$_2$O$_3$/YSZ (111) used for calibration. (a) RHEED specular intensity of a NiO film during growth.  (B) Fourier transform of the specular intensity showing a sharp peak at the monolayer frequency. (c-d) Post-growth RHEED images of the NiO film. (e-h) Same for a representative Nd$_2$O$_3$ film.}
\end{figure}

\subsection{RHEED Oscillation Calibration}
Initial flux measurements are obtained by monitoring RHEED oscillations during the growths of Nd$_2$O$_3$ on (ZrO$_2$)$_{0.905}$(Y$_2$O$_3$)$_{0.095}$ (111) and NiO on MgO (100) using the parameters outlined in Sun et al. \cite{Sun2022}.  In \hyperref[S:OSC]{Fig. \ref*{S:OSC}} we detail the RHEED oscillation measurements of NiO and Nd$_2$O$_3$ in (a-d) and (e-h), respectively.  While both compounds grow over a very wide range of conditions (ozone flux, substrate temperature) we find the optimal conditions for RHEED oscillations of NiO are around 450 $^{\circ}$C~ (following an 800 $^{\circ}$C~ vacuum anneal to clean the MgO surface).  Conversely, it is preferable to grow Nd$_2$O$_3$ at higher temperatures, $\sim 970$~$^{\circ}$C to stabilize the hexagonal structure; though often the first few layers will take on a bixbyite structure, evidenced by a noticeably different oscillation frequency as seen in (e).  By monitoring the RHEED specular intensity in (a) and (e) and performing a Fourier transform, (b) and (f), the monolayer formation time can be easily extracted and converted into an elemental flux.   

\begin{figure}
  \resizebox{12.5 cm}{!}{
  \includegraphics{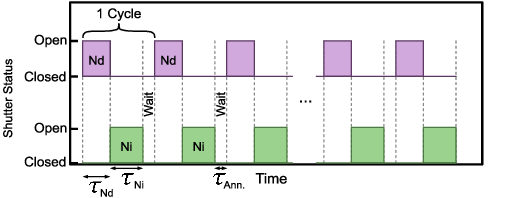}}
  \caption{\label{S:shutter} Shutter timing diagram illustrating the sequential deposition of NdNiO$_3$ thin films. The shutters for the neodymium and nickel sources are opened in sequence for time periods $\tau_{\textrm{Nd}}$ and $\tau_{\textrm{Ni}}$, respectively.  Between each cycle the sample is ozone annealed with the shutters closed for a short period of $\tau_{\textrm{Ann.}}=5$~seconds to promote full oxidation of the nickel layer, as in Ref. \cite{King2014}.  }
\end{figure}

\begin{figure}
  \resizebox{14.5 cm}{!}{
  \includegraphics{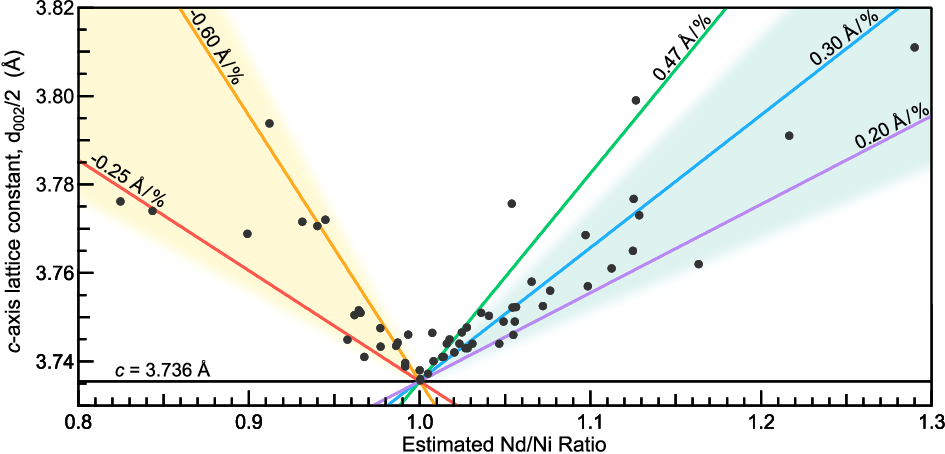}}
  \caption{\label{S:XRD} X-ray diffraction calibration data for 20 u.c. thick films of NdNiO$_3$ on (001) SrTiO$_3$. $c$-axis lattice constant, determined from the 002 peak position, as a function of the nominal cation ratio Nd:Ni. Slopes of typical correction factors employed in the optimization procedure are labeled.}
\end{figure}

\newpage
\subsection{XRD and XRR Calibration}
While the aforementioned calibration method greatly improves upon the basic calibration using a QCM, we find that its accuracy is typically limited to the the few percent scale for each source and as such requires further refinement.  This is accomplished using the method described by Li et al. in Ref. \onlinecite{Li2021}:  the $c$-axis lattice constant (in our case the 002 peak position) is used as a proxy for non-stoichiometery with the minimal value of the lattice constant corresponding to a stoichiometric cation ratio of Nd:Ni = 1:1.  Our repetition of this procedure produces much the same qualitative characteristics as in Ref. \cite{Li2021}, although with a slightly lower minimal lattice constant of $c=3.736$~\AA, and is detailed in \hyperref[S:XRD]{Fig. \ref*{S:XRD}}.  To construct this diagram we start by growing a sequence of films with differing shutter times, $\tau_{\textrm{Nd,Ni}}$, and record the ratio $\nicefrac{\tau_{\textrm{Nd}}}{\tau_{\textrm{Ni}}}$.  The shutter times $\tau_{\textrm{Ndi}}$~ and $\tau_{\textrm{Ni}}$~ correspond to our best estimate of the time to deposit a single monolayer of NdO and NiO$_2$, respectively. A timing diagram showing the sequential opening of the Nd and Ni shutters in a constant background partial pressure of O$_3$ with a short, 5 second, anneal between cycles is provided in \hyperref[S:shutter]{Fig. \ref*{S:shutter}}.  While the flux ratio $\nicefrac{\Phi_{Nd}}{\Phi_{Ni}}$ is not known \textit{a priori}, once the growth sequence converges to near the minimal value this ratio can be calculated from
\begin{equation*}
    1 = \frac{\Phi_{\textrm{Nd}}}{\Phi_{\textrm{Ni}}} \times \frac{\tau_{\textrm{Nd,f}}}{\tau_{\textrm{Ni,f}}}.
\end{equation*}
and applied to the whole sequence to determine the Nd:Ni ratio for all films.  \hyperref[S:XRD]{Figure \ref*{S:XRD}} is an aggregation of many such growth runs.  The scatter increases as we depart from Nd:Ni = 1:1, which is a consequence of source drift on the above procedure.  Though modern effusion cells provide excellent flux stability, often reducing flux drift to less then 1\% per hour, the difference between $\nicefrac{\tau_{\textrm{Nd}}}{\tau_{\textrm{Ni}}}$ at the start of the run and the end (when 1:1 is reached) means that estimation of Nd:Ni for the first samples in the sequence (Nd:Ni $\gg 1$) is less accurate than at the end (Nd:Ni $\sim 1$).  This estimated cone of uncertainty is illustrated by the shaded regions.  Despite this uncertainty in the non-stoichiometry of films well away from 1:1, in the region near the minimum lattice constant discernible changes are still observed down to single percent changes in the shutter times, evidencing that this method can be used to hone the non-stoichiometry down to the percent level or less.

A typical optimization routine at the start of a growth campaign consists of starting with the RHEED oscillation calibration of one or both of Nd and Ni. Then the starting shutter times are calculated to begin with a few percent of excess Neodymium to bias the initial growth towards Nd:Ni $> 1$; this is to ensure that the direction of the first step is known \textit{a priori} (\textit{i.e.}, reducing the Nd shutter time) and to avoid venturing into the Ni-rich side of the diagram.  We find that on the Ni-rich side the relationship between $d_{002}$ and the shutter times is less repeatable.  We attribute this to the propensity of excess Ni to precipitate out of the film in the form of NiO, rather then readily incorporating like neodymium does.  The flux ratio can then be `walked-in' to 1:1 by progressive adjustment of the shutter times, growth, and XRD measurement.  If the starting Nd excess is greater then 10\% or so an `aggressive' adjustment factor of 0.2 to 0.3 \AA/\% can be used in tuning the shutter times between subsequent films; as Nd excess is further reduced more conservative values of 0.3 to 0.47 \AA/\% are used to avoid accidentally jumping over onto the Ni-rich side.  Once the Nd:Ni ratio is within 5-7\% of the 1:1 stoichiometric ratio, the total film thickness can also be fine-tuned (by changing $\tau_{\textrm{Nd}}$ and $\tau_{\textrm{Ni}}$ by the same fraction) to match the desired number of unit cells, typically 20, based on x-ray reflectivity (XRR) measurements of the NdNiO$_3$ film thickness.

\subsection{Effects of Non-Stoichiometry on Transport}
\begin{figure}
  \resizebox{11 cm}{!}{
  \includegraphics{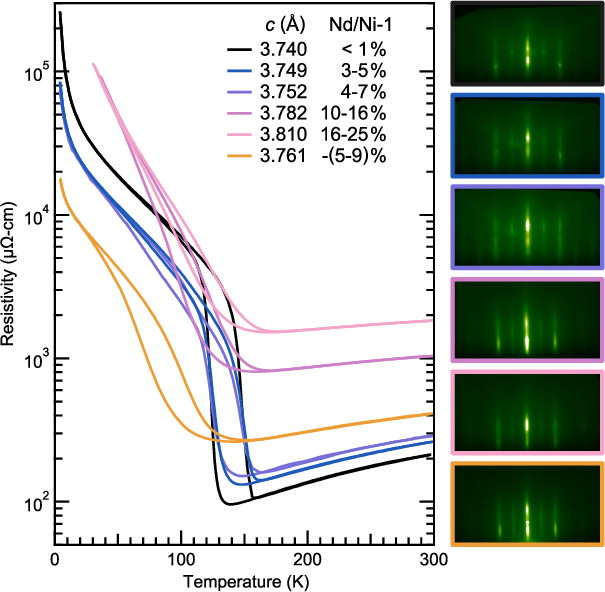}}
  \caption{\label{S:cations} Effects of cation nhon-stoichiometry on 20 u.c. thick NdNiO$_3$ thin films grown on (001)-oriented SrTiO$_3$. Resistivity measurements of films of varying non-stoichiometry from slightly nickel rich to highly neodymium rich.  Post-growth RHEED images, viewed along the [110]$_{\textrm{pc}}$ azimuth, of each of the films are pictured on the right. All images use a linear intensity scale. }
\end{figure}

Here we describe the effects of non-stoichiometry on the transport properties of NdNiO$_3$ films grown on (001) SrTiO$_3$.  The temperature dependent resistivity of a sequence of samples of varying lattice constants (and estimated non-stoichiometry based on \hyperref[S:XRD]{Fig. \ref*{S:XRD}}) are presented in \hyperref[S:cations]{Fig. \ref*{S:cations}}.  The film with the lowest lattice constant has a low room-temperature resistivity of $\sim 200 \mu\Omega$-cm, and a sharp metal-to-insulator transition (MIT) with a jump of nearly two decades.  The structural quality is also evident in the accompanying RHEED image, where sharp diffraction peaks are visible on the primary and half-order streaks.  As the Nd:Ni ratio is increased a clear increase in the resistivity is visible (up to an order of magnitude), along with a broadening of the MIT and widening of the hysteresis loop. At the same time the sharp diffraction peaks fade and the RHEED streaks become `peanut shaped' -- likely due to an increase in the film roughness.  We note a marked difference between Nd-rich films under tensile strain when compared to prior measurements under compressive strain \cite{Breckenfeld2014a,Preziosi2017} where excess Nd drives the films metallic.  As previously noted in studies of non-stoichiometry on SrTiO$_3$ \cite{Yamanaka2019a} the MIT temperature \textit{increases} with increasing Nd content, and the MIT is not suppressed even up to very high levels (Nd:Ni $\sim 1.25$).  Finally, on the nickel-rich side of the diagram broadening of the MIT and an overall increase of the resistivity are observed.  These results indicate that a stoichiometric film on SrTiO$_3$ is the one with a minimum lattice constant, maximum transition sharpness, and minimum resistivity of the metallic state.

\subsection{Additional Aging Effect Data}
\begin{figure}
  \resizebox{11 cm}{!}{
  \includegraphics{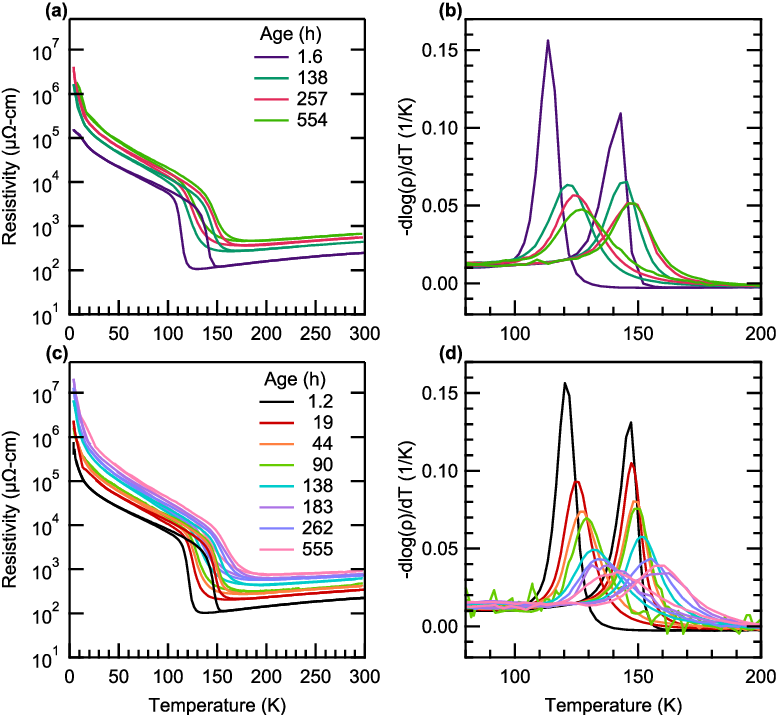}}
  \caption{\label{S:aging} Aging effects on the transport properties of stoichiometric NdNiO$_3$ films grown on SrTiO$_3$. (a) Resistivity of a single NdNiO$_3$ film measured periodically following growth, reproduced from the main text. (b) Logarithmic derivative of the transport curves in panel (a). (c) Resistivity of another film measured at a higher frequency over the same period of time as the one pictured in panel (a). (d) Logarithmic derivative of the resistivity curves in (c).}
\end{figure}

In \hyperref[S:aging]{Fig. \ref*{S:aging}} (unshifted) raw data corresponding to the sample aging study presented in Fig. 2 of the main text.  \hyperref[S:aging]{Figs. \ref*{S:aging}(a)} and \hyperref[S:aging]{\ref*{S:aging}(b)} are reproduced from the main text for reference and \hyperref[S:aging]{Figs. \ref*{S:aging}(c)} and \hyperref[S:aging]{\ref*{S:aging}(d)} show data for the sample that was measured at much shorter intervals following growth.  It is evident that in both cases a gradual increase of the film resistivity is present over time, with the room temperature values increasing by a factor of 4-5 over the course of  $\sim 3.5$~ weeks while stored in a dry, N$_2$ purged desiccator.

\clearpage
\section{Infinite-Layer Nickelates}

\subsection{AFM and XRD of Optimized Films}
\begin{figure}[ht]\centering
  \resizebox{\columnwidth}{!}{\includegraphics{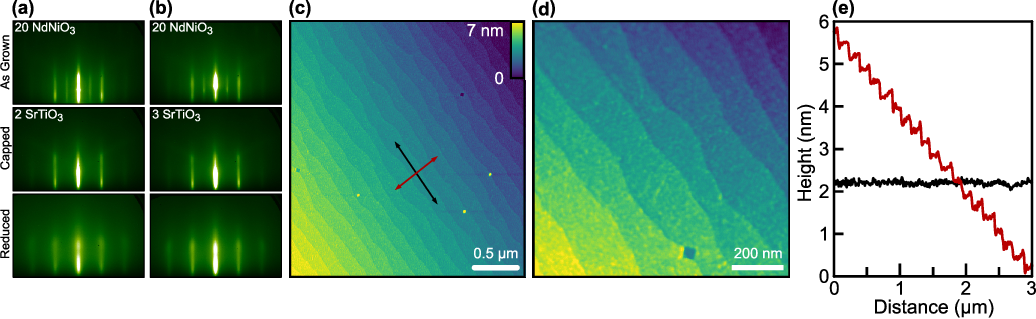}}
  \caption{\label{S:AFM} Measurements of the surface quality of NdNiO$_2$ films reduced using atomic hydrogen. (a) RHEED images taken along the [110]$_{pc}$~azimuth of an as grown 20 u.c. thick NdNiO$_3$~film, the same film after capping with 2 u.c. of SrTiO$_3$, and after atomic hydrogen reduction. (b) The same as in (a) but with a 3 u.c. thick SrTiO$_3$ cap.  (c) AFM image of a 20 u.c. thick NdNiO$_2$ film capped with 2 u.c. of SrTiO$_3$ and reduced with atomic hydrogen; the extracted terrace-averaged RMS roughness is $124 \pm 10$~pm. (d) Zoom in on the stepped terraces. (e) Line profiles along (black) and across (red) the terraces showing preservation of the substrate's atomic steps.  RHEED images in (a) and (b) use a linear intensity scale.}
\end{figure}

\begin{figure}[ht]\centering
  \resizebox{0.75\columnwidth}{!}{\includegraphics{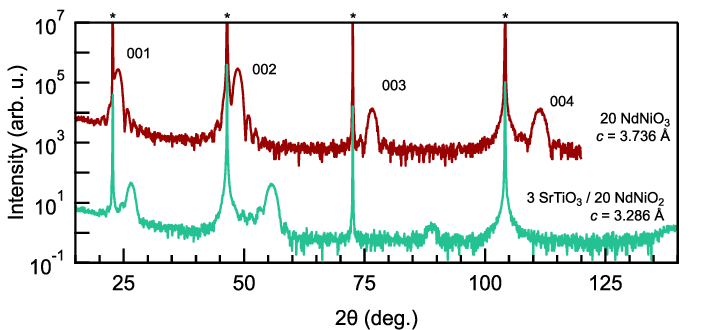}}
  \caption{\label{S:optXRD} Diffraction intensity along the specular direction for an uncapped 20 u.c. thick NdNiO$_3$ film and a 3 u.c. SrTiO$_3$ capped NdNiO$_2$ film grown on SrTiO$_3$ showing the $00L$ reflections. Nelson-Riley analysis of the infinite-layer film peaks gives a lattice constant of $c=3.286$~\AA. (*) indicates reflections from the (001) SrTiO$_3$ substrate.}
\end{figure}

In this section we present additional measurements of optimized, reduced NdNiO$_2$ films on SrTiO$_3$. In \hyperref[S:AFM]{Fig. \ref*{S:AFM}} we show RHEED images and AFM topography of a reduced infinite-layer films. \hyperref[S:AFM]{Figs. \ref*{S:AFM}(a)} and \hyperref[S:AFM]{\ref*{S:AFM}(b)} present RHEED images of additional 20 u.c. thick NdNiO$_3$ films after growth, after capping, and after atomic hydrogen reduction. In both reduced films, reduced using the optimized three-step procedure, the preservation of the RHEED streaks indicates the film surface remains ordered. The RHEED images of both films are qualitatively similar, though the 3 u.c. capped film appears to have a slightly lower incoherent background then that of the film with the thinner cap.  \hyperref[S:AFM]{Figures \ref*{S:AFM}(c)-(e)} show AFM measurements of a 20 u.c. thick NdNiO$_2$ film capped with 2 u.c. of SrTiO$_3$ and reduced with atomic hydrogen.  The reduced film shows a stepped terrace structure analogous to the atomically flat substrate on which it was grown, with a low surface roughness of $\sim 124$~pm.  This is similar to films reduced using the solid-state aluminum reduction process introduced in Ref. \onlinecite{Wei2023}.  In \hyperref[S:optXRD]{Fig. \ref*{S:optXRD}} we show larger area $\theta-2\theta$ scans of a perovskite, NdNiO$_3$, film and an optimized, reduced infinite-layer, NdNiO$_2$, film which cover all of the $00L$ peaks observable with Cu K$\alpha$ x-rays.  In the infinite-layer film, all four reflections are visible, with the $004$ peak sitting at the end of the range accessible with our goniometer.

\begin{figure}[ht]\centering
  \resizebox{0.80\columnwidth}{!}{\includegraphics{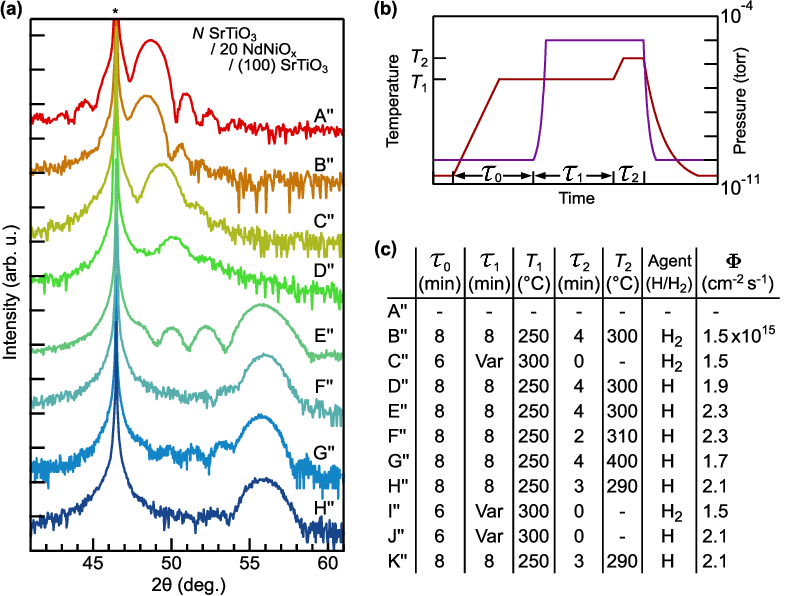}}
  \caption{\label{S:conditions} Conditions used for H and H$_2$ reduction of films during experiments exploring the effect of the SrTiO$_3$ cap on reduction. (a) Replica of Fig. 7(a) from the main text with samples labeled. (b) Cartoon of a generic three-step ($\tau _2 \neq 0$) or two-step ($\tau_2 = 0$) reduction. (c) Table of reduction conditions used in the preparation of samples in (a) as well as in \hyperref[S:STOH2]{Figs. \ref*{S:STOH2}}, \hyperref[S:uncappedH]{\ref*{S:uncappedH}}, and \hyperref[S:amorphous]{\ref*{S:amorphous}}. }
\end{figure}

\subsection{Reduction Effects on the Film Surface}
In this section we present additional data concerning the effects of molecular and atomic hydrogen on the film surface.  In \hyperref[S:conditions]{Fig. \ref*{S:conditions}} we present the details of the reduction program used for samples in Fig. 7 (samples A''-G'') of the main text as well as those presented later in this section (samples I'',J'').  Of these, several samples were subjected to several sequential single-step reduction cycles ($\tau_2 = 0$) with variable times, performing RHEED measurements between cycles. These include sample C'', which was reduced in four increments of 15, 15, 15, and 45 minutes (90 min total), sample I'' in four increments of 15, 30, 30, and 15 minutes (90 min total), and J'' in three increments of 15, 15, and 30 minutes (60 min total).  Fluxes are quoted in atoms/cm$^2$/sec for samples reduced with atomic hydrogen  and in molecules/cm$^2$/sec for those reduced with molecular hydrogen only. In all cases a consistent hydrogen flow rate of 0.72 sccm was used (giving a chamber pressure of $\sim 9.5\times 10^{-6}$~ torr) and the atomic hydrogen flux was adjusted by tuning the cracker temperature.

\begin{figure}[ht]\centering
  \resizebox{\columnwidth}{!}{\includegraphics{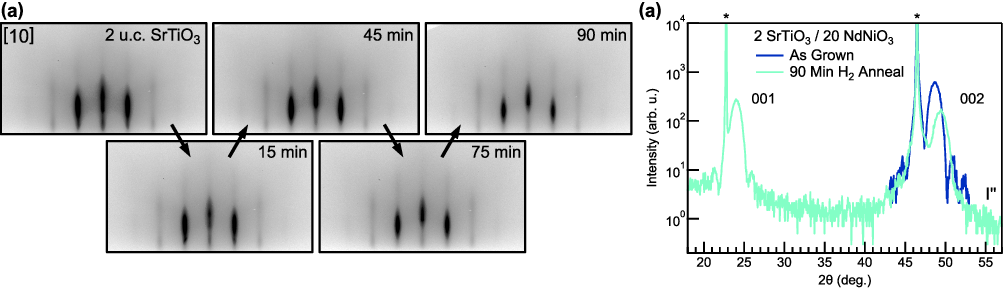}}
  \caption{\label{S:STOH2} Molecular hydrogen reduction of a 20 u.c. thick NdNiO$_3$ / SrTiO$_3$ film capped with 2 u.c. of SrTiO$_3$. (a) RHEED images along the [10] azimuth of a film exposed to H$_2$, at a temperature of 300 $^{\circ}$C, for repeated intervals accumulating to 90 minutes total. (b) XRD measurements of the same film before and after the 90 minutes of total reduction time. (*) indicates reflections from the (001) oriented SrTiO$_3$ substrate.}
\end{figure}

\begin{figure}[ht]\centering
  \resizebox{\columnwidth}{!}{\includegraphics{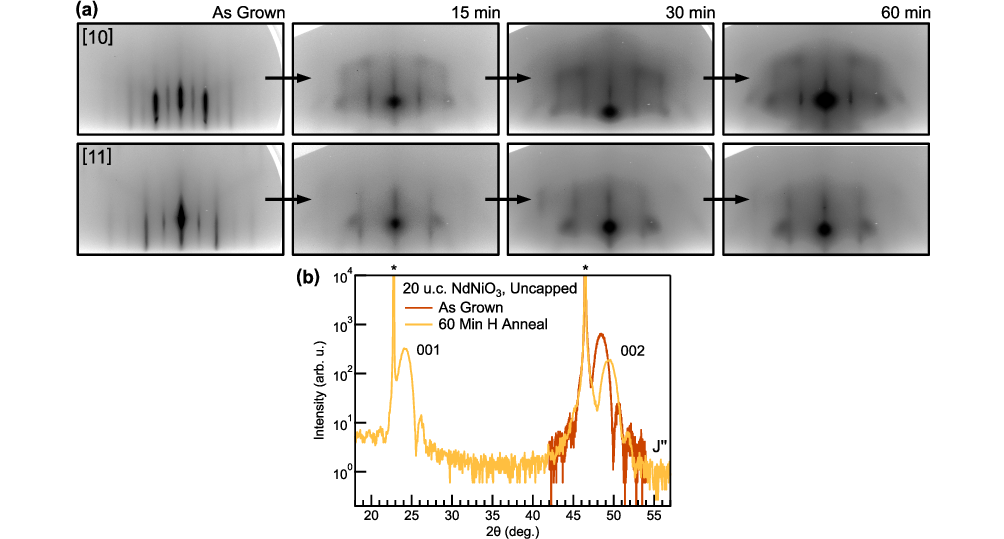}}
  \caption{\label{S:uncappedH}  Effects of atomic hydrogen reduction on an uncapped, 20 u.c. thick NdNiO$_3$ / SrTiO$_3$ film.  (a) RHEED images along the [100]$_{\textrm{pc}}$ and [110]$_{\textrm{pc}}$ azimuths of a uncapped film; images are taken of the as grown film surface as well as after 15, 30, and 60 minutes of atomic hydrogen exposure time at a sample temperature of 300 $^{\circ}$C. (b) XRD measurements of the same film before and after the 60 minute exposure. (*) indicates reflections from the (001) oriented SrTiO$_3$ substrate.}
\end{figure}

In \hyperref[S:STOH2]{Figs. \ref*{S:STOH2}} and \ref{S:uncappedH} we detail two additional progressive annealing experiments.  \hyperref[S:STOH2]{Figure \ref*{S:STOH2}} shows a sample with a 2 u.c. thick SrTiO$_3$ cap annealed in molecular hydrogen.  As can be seen in the RHEED images, the molecular hydrogen has little-to-no impact on the SrTiO$_3$ surface, even for long exposure times.  Similar to the uncapped sample shown in the main text, molecular hydrogen is insufficient to produce the infinite-layer phase (under these conditions) and instead an oxygen-deficient perovskite phase is visible in the XRD, with out-of-plane lattice paramter $c\sim 3.69$~\AA, following 90 minutes of exposure.  \hyperref[S:uncappedH]{Figure \ref*{S:uncappedH}} shows a sample that was exposed to atomic hydrogen in progressive steps.  Similar to the sample presented in the main text, the RHEED images indicate a rapid degradation of the film surface and the presence of other phases (faint rings and additional spots).  Even after a relatively long exposure, 60 minutes, the majority of the film remains an oxygen-deficient perovskite ($c\sim 3.69$~\AA) without any indication of formation of the infinite-layer phase, evidenced by the XRD patterns in \hyperref[S:uncappedH]{Fig. \ref*{S:uncappedH}(b)}.

\begin{figure}[ht]\centering
  \resizebox{0.9\columnwidth}{!}{\includegraphics{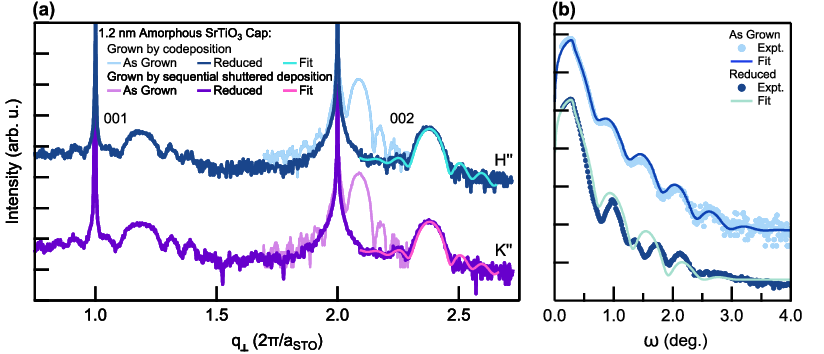}}
  \caption{\label{S:amorphous}  Atomic hydrogen reduction of 20 u.c. thick NdNiO$_3$ / SrTiO$_3$ films capped with 1.2 nm of amorphous SrTiO$_3$.  (a) XRD of nickelate films as grown (light) and after capping and reduction (dark); capping layers were grown using either simultaneous codeposition of Sr and Ti (blue) or sequential shuttered deposition (purple) at room temperature. 002 peaks of the reduced films are fit with an $N$-slit interference pattern (teal, pink) to extract the coherent thickness. (b) Low-angle XRR of one of the samples before and after capping/reduction (dots) and corresponding best fits (lines) using a three slab model.}
\end{figure}

Finally, we discuss the effects of amorphous capping layers on the reduction of 20 u.c. NdNiO$_3$/SrTiO$_3$ films using atomic hydrogen.  X-ray diffraction measurements showing the 001 and 002 peak regions for two films with amorphous caps are provided in \hyperref[S:amorphous]{Fig. \ref*{S:amorphous}(a)}.  For both films, the capping layer was deposited in a background pressure of $2\times 10^{-6}$~ torr of distilled O$_3$~ with the sample at room temperature.  For sample H'' the capping layer was deposited with Sr and Ti shutters opened simultaneously (codeposition) and for sample K'' the shutters were opened sequentially (shuttered deposition); in both cases no streaks, spots, or rings were visible in RHEED after deposition of the SrTiO$_3$ layers.  The thickness of both capping layers was 1.2 nm -- equivalent to 3 u.c. of SrTiO$_3$~ -- and the subsequent reductions, using the same conditions, appear to progress similarly based on the XRD measurements in \hyperref[S:amorphous]{Fig. \ref*{S:amorphous}(a)}.  The obvious difference in XRD patterns between those samples reduced (using atomic hydrogen) with an amorphous cap (H'' and K'') versus those reduced with no cap (D'' and J'') illustrate the importance of the overlayer in facilitating the reduction even in the absence of epitaxial stabilization.  

We do note, however, that the quality of the amorphous SrTiO$_3$~ capped samples are somewhat reduced compared to those with a crystalline cap.  Fitting of the XRD patterns of both the reduced films with an $N$-slit interference pattern,
\begin{equation*}
    I \propto \frac{1}{N}\frac{\sin^2(Naq/2)}{\sin^2(aq/2)},
\end{equation*}
gives a Scherrer thickness of only 4.43 nm, \textit{i.e.} $N\sim 13.5$, indicating that the coherent crystalline region does not extend across the entirety of the film thickness. Additionally, the x-ray reflectivity, \hyperref[S:amorphous]{Fig. \ref*{S:amorphous}(b)}, of the reduced film cannot be accurately fit using a three-slab model (SrTiO$_3$/NdNiO$_3$/SrTiO$_3$) as was done for the crystalline film in Fig. 5(d) of the main text. The best-fit (green) for this reflectivity curve gives a roughness for the NdNiO$_2$ layer, 3.2 nm (compared to that of the as grown film, 0.48 nm), and a thickness of 5.6 nm (intermediate to the Scherrer thickness of 4.4 nm and the expected thickness of 6.5 nm).  These results are consistent with the capping layer providing some measure of epitaxial support during the reduction process, which is absent in those films with an amorphous cap, and a fraction of the film either forms a defect phase or decomposes -- as reported previously in CaH$_2$ reduced films \cite{Lee2020}.  We cannot rule out, however, that the observed differences between the crystalline and amorphous SrTiO$_3$ capped films result from differences in transport of the reactants (\textit{e.g.} H or O$_2$) through the different matrices.  We emphasize, however, that the role of the capping layer, at least in atomic hydrogen reduction experiments, clearly goes beyond that of epitaxial stabilization.  In the XRD patterns of uncapped films, D'' and J'', the effect of the reduction is \textit{not} to cause decomposition of the entire thickness of the film, but rather to form a polycrystalline scale on the surface and to hamper further reduction -- as evidenced by the minor 002 peak shifts observed in XRD.  This is in contrast with the more modest observed differences between crystalline capped (Samples F'', G'') and amorphous capped (Samples H'', K'') films which all exhibit similar 002 peak positions.

\clearpage
\bibliography{Nickelates}